\documentstyle[12pt]{article}

\newlength{\dinwidth}
\newlength{\dinmargin}
\def\cN{{\cal N}}
\def\Dzero{D${0}$}
\def\GeV{{\rm GeV}}

\def\lapproxeq{\lower .7ex\hbox{$\;\stackrel{\textstyle
<}{\sim}\;$}}
\def\gapproxeq{\lower .7ex\hbox{$\;\stackrel{\textstyle
>}{\sim}\;$}}

\begin{document}
\titlepage
\begin{flushright}
DTP/96/44  \\
RAL--TR--96--037 \\
May 1996 \\
\end{flushright}

\begin{center}
\vspace*{2cm}
{\Large \bf Parton distributions: a study of the new\\[2mm]
HERA data, $\alpha_s$, the gluon and $p \bar p$ jet production} \\
\vspace*{1cm}
 A.\ D.\ Martin$^a$, R.\ G.\ Roberts$^b$
and W.\ J.\ Stirling$^{a,c}$ \\

\vspace*{0.5cm}
$^a \; $ {\it Department of Physics, University of Durham,
Durham, DH1 3LE }\\

$^b \; $ {\it Rutherford Appleton Laboratory, Chilton,
Didcot, Oxon, OX11 0QX}\\

$^c \; $ {\it Department of Mathematical Sciences, University of Durham,
Durham, DH1 3LE }
\end{center}

\vspace*{4cm}
\begin{abstract}
New data, especially HERA measurements of the proton structure
function at small $x$, allow the opportunity to improve our knowledge
of the gluon and quark distribution functions.
We perform a global analysis which incorporates these new precise data,
and which extends down to $Q^2 = 1.5\; \GeV^2$. We discuss the sensitivity
to the value of $\alpha_s$ and the improvement 
in the determination of the gluon. We compare the predictions of the single
jet inclusive cross section with recent measurements from the Fermilab 
Tevatron. 
\end{abstract}

\newpage

Predictions for hard scattering processes involving hadrons rely on 
a precise knowledge of the parton distribution functions. Traditionally these
are determined from global analyses of data for a wide range of processes,
see for example \cite{mrsa,cteq}. As the precision of data improves and
the kinematic domain of the measurements enlarges we obtain a more
rigorous test of perturbative QCD and a better knowledge of the parton
distributions. This is well illustrated by the new measurements 
\cite{h1,zeus} of the 
proton structure function at HERA, which are more precise 
and extend to lower $x$ than hitherto.

The significant improvement in the data demands a concomitant
refinement of the parton distributions. In this Letter we 
describe a new global analysis, based on next-to-leading-order
DGLAP evolution, of the type described in Ref.~\cite{mrsa}
but with the following special features:
\begin{itemize}
\item[(i)] structure function data down to $Q^2 = 1.5\; \GeV^2$
 are included;
\item[(ii)] fits are performed  for two different
values of the QCD coupling, the standard
``deep inelastic scattering" value ($\alpha_s(M_Z^2) = 0.113$) \cite{mv}
and a larger value ($\alpha_s(M_Z^2) = 0.120$) suggested, for example,
 by measurements at LEP \cite{siggi};
\item[(iii)] particular attention is paid to the
small $x$ behaviour of the gluon
and sea quark distributions;
\item[(iv)] the single jet inclusive cross section is predicted and compared
with the new CDF \cite{cdfjet} and \Dzero\ \cite{d0jet} measurements at the
Fermilab $p \bar p$ collider.
\end{itemize}
We elaborate on these aspects of the analysis in turn below.

In previous analyses we fitted to structure function data with
$Q^2 > 5 \; \GeV^2$.\footnote{Contamination by higher-twist
contributions at large $x$ is avoided by the additional cut
 $W^2 = Q^2(1-x)/x > 10\; \GeV^2$.}
However the ``dynamical" partons of the GRV model \cite{grv}
were found to give a good description of data down 
to $Q^2 \sim 1\; \GeV^2$, although
the new HERA measurements \cite{h1,zeus} at low $Q^2$
do show some systematic discrepancy between the GRV predictions and the data.
In this respect it is important to investigate whether this signals a
problem for DGLAP evolution at such low scales
or for the assumptions implicit in the dynamical parton model.
We therefore include all data with $Q^2 \geq 1.5\; \GeV^2$.
The choice $\alpha_s(M_Z^2) = 0.113$  corresponds to the optimum QCD
coupling as determined by the scaling violations of the fixed-target
deep inelastic data. The key data here are the BCDMS $F_2^{\mu p,d}$
measurements \cite{bcdms} in the interval $0.3 \lapproxeq x \lapproxeq
0.5$. There is, however, some indication that the HERA structure
function data and the Fermilab jet data prefer a larger value of the
coupling \cite{mrsv,bf,ggy,mrsjet}.
To illustrate this point we therefore also present  results
of a second global fit in which the coupling is fixed at
$\alpha_s(M_Z^2) = 0.120$. We call the resulting set of
partons\footnote{These new distributions were presented in preliminary
form at the Rome DIS96 Workshop.} R$_1$ and R$_2$ respectively.

The advent of the HERA measurements of $F_2$ has considerably improved
our knowledge of the gluon and sea quark distributions  in the small $x$
regime. We may represent their leading small $x$ behaviour  by the forms
\begin{equation}
xg \to x^{-\lambda_g(Q^2)}\; , \qquad xS \to x^{-\lambda_S(Q^2)} \; .
\label{lambdadef}
\end{equation}
In the perturbative region the evolution of the sea quark distributions
is driven by the gluon, which is the dominant parton at small $x$, via
the $g \to q \bar q$ splitting. The distributions are therefore
correlated at small $x$, and for this reason most previous global
analyses set $\lambda_S = \lambda_g$ in (\ref{lambdadef}) at the input
scale, usually taken to be $Q_0^2 = 4\; \GeV^2$. Even disregarding
non-perturbative effects, this expectation is only approximate. As
we evolve up in $Q^2$ using the DGLAP evolution equations,
$\partial g / \partial\ln Q^2 = P_{gg} \otimes g + \ldots$ and
$\partial q / \partial\ln Q^2 = P_{qg} \otimes g + \ldots$, the effective
exponents $\lambda_i(Q^2)$ increase with $Q^2$, and satisfy
\begin{equation}
\lambda_S(Q^2) = \lambda_g(Q^2) - \epsilon \; ,
\label{eq:dglap}
\end{equation}
where $\epsilon$ is positive and slowly varying with $Q^2$, provided
that the evolution is sufficiently long. Typically we find
 $\epsilon \sim 0.05$
at $Q^2 \sim 200\; \GeV^2$, see below. That is, DGLAP evolution leads to a
sea distribution which is slightly less steep in $x$ than the gluon
distribution.

The precision of the new HERA small $x$ structure function data is now such that
we can begin to study the small $x$ behaviour of the quarks and gluons
in detail. We therefore allow both $\lambda_S$ and $\lambda_g$ to be free
parameters at the input scale, which we here take to be $Q_0^2 = 1\;
\GeV^2$. To see the extent to which $\lambda_S$ and $\lambda_g$ can  be
individually determined, we also perform fits with $\lambda_S =
\lambda_g$ at $Q_0^2$.  Where appropriate, we show properties of these
additional parton sets, which we label R$_3$ and R$_4$ corresponding
to $\alpha_s(M_Z^2)  = 0.113$ and $0.120$ respectively.

Another interesting development  is the constraint that the high
statistics jet data from Fermilab are starting to impose on the partons
and the QCD coupling $\alpha_s$. The slope of the inclusive jet
transverse energy ($E_T$) distribution is particularly sensitive to the
value of $\alpha_s$ \cite{mrsv,ggy}. In the region where the data are
most definitive, $50 \lapproxeq E_T \lapproxeq 200\; \GeV$, the jets
arise from gluon-initiated QCD processes. Hence the data can give a
tight constraint  on $\alpha_s(\mu^2) g(x,\mu^2)$ where the gluon is
sampled in the region $x \sim 2E_T/\sqrt{s} \sim 0.1$, but at a high
scale $\mu^2 \sim E_T^2$ where evolution tends to wash out the
differences between the parton sets.  We compare the predictions based
on our new R$_1$ and R$_2$ sets with CDF \cite{cdfjet} and preliminary
\Dzero\ \cite{d0jet} jet $E_T$ distributions to demonstrate the potential
discriminating power of the data.

As in our previous studies, we adopt the following simple
analytic form for the starting distributions:\footnote{The
distributions are defined in the $\overline{{\rm MS}}$ renormalization and 
factorization schemes.}
\begin{equation}
xf_i(x,Q_0^2)  = 
 A_i x^{-\lambda_i}(1-x)^{\eta_i}(1 + \epsilon_i \sqrt{x} + \gamma_i x)\; , 
\label{eq:starting}
\end{equation}
for $i=u_V,\; d_V,\; S,\; g$, but at the lower scale $Q_0^2 =
1$~GeV$^2$.
Here $S$ represents the total  sea quark distribution. At $Q_0^2$,
only the $u,d,s$ distributions are non-zero, and so
$S \equiv  2(\bar{u}+\bar{d}+\bar{s})$.
The flavour structure of the sea is taken to be 
\begin{eqnarray} 
2\bar{u} & = & 0.4  S - \Delta \nonumber \\ 
2\bar{d} & = & 0.4  S + \Delta \nonumber  \\ 
2\bar{s} & = & 0.2  S \; ,
\label{eq:sea} 
\end{eqnarray}  
with 
\begin{equation} 
x\Delta \; \equiv \; x(\bar{d}-\bar{u}) \; = \; A_{\Delta} x^{0.3} 
(1-x)^{\eta_S} (1 + \gamma_{\Delta}x) . 
\label{eq:delta} 
\end{equation} 
The suppression of the strange distribution is motivated by data on
 neutrino-induced deep inelastic dimuon production  obtained 
 by the CCFR collaboration \cite{CCFR2}. In our previous studies this
factor 2 suppression was applied at $Q_0^2 = 4$~GeV$^2$, rather than
at $Q_0^2 = 1$~GeV$^2$ as in (\ref{eq:sea}). The consequence is that the
strange sea is slightly larger at $Q^2 = 4$~GeV$^2$ and, in fact, in
better agreement with the CCFR measurement. That is, the strange sea
lies more centrally in the allowed band for $s(x,Q^2 = 4\; {\rm
GeV}^2)$, determined by the next-to-leading order CCFR analysis, than
does the MRS(A) curve shown in Fig.~4 of Ref.~\cite{mrsa}.
The $u,d$ flavour symmetry breaking
distribution $\Delta(x)$ is chosen to give good agreement with the NMC
measurement of the Gottfried sum \cite{NMC}, and the NA51 measurement
of the  Drell-Yan $pp/pn$ asymmetry \cite{NA51}. 

In previous analyses, we treated the charm quark as massless and evolved
its distribution from zero at $Q^2 = m^2$, where $m^2 = 2.7$~GeV$^2$ was
chosen  to reproduce the EMC charm data \cite{emc}. Now that we
have lowered the starting scale to $Q_0^2 = 1$~GeV$^2$ it is necessary
to smooth out the onset of charm. We take $c(x,Q_0^2) = 0$ and evolve up
in $Q^2$ with four flavours but since $Q_0^2$ is now much lower, this
unmodified density would overestimate the observed $F_2^c(x,Q^2)$.
We simply suppress the generated charm density by a smooth factor \cite{gp}
which mimics the threshold behaviour of the massive quark, i.e. 
\begin{equation}
c(x,Q^2) \longrightarrow \left\{ 1 - \cN\left(\frac{m_0^2}{Q^2}
\right) \right\}\; c(x,Q^2) 
\end{equation}
where
\begin{equation}
\cN(z) = 6z\; \left[ 1 - \frac{2z}{\sqrt{1+4z}}\ln\left( 
{ \sqrt{4z+1} + 1 \over \sqrt{4z+1} - 1 }
\right)\right]
\end{equation}
with $m_0^2=3.5$ GeV$^2$ chosen give a reasonable description of the
EMC \cite{emc} and preliminary H1 \cite{h1charm} measurements of the charm
structure function $F_2^c$. We see from Fig.~1 that this approximate
treatment of charm does indeed give a satisfactory description of $F_2^c$ at
present, but as the data improve it will become necessary to 
give a more satisfactory treatment of charm-mass effects.

Using the above parton parametrizations at $Q_0^2 = 1$~GeV$^2$ we
perform global next-to-leading order DGLAP fits to the deep inelastic
and related hard scattering data that were used in the MRS(A) analysis
\cite{mrsa}, but now including data down to $Q^2 = 1.5$~GeV$^2$,
supplemented by  SLAC \cite{slac}, E665 \cite{e665} and
updated NMC \cite{nmcf2} structure
function measurements and, most importantly, the new H1 \cite{h1} and
ZEUS \cite{zeus} $F_2$ data. We present the results of the four
different fits R$_1$ -- R$_4$, which we described above. The optimum
values of the starting distribution parameters
are listed in Table~\ref{tab:paras}.
Note that three of the four $A_i$ 
coefficients are determined by the momentum and flavour sum rules. The
fourth ($A_S$) is a fitted parameter. Because of the different values
of $Q_0^2$, these parameters cannot be directly compared with the
corresponding parameters from previous MRS analyses. 
\begin{table}[tbh]
\begin{center}
\begin{tabular}{|lcrrrr|}
\hline
\rule[-1.2ex]{0mm}{4ex} &  & R$_1$~ &R$_2$~ & R$_3$~ & R$_4$~ \\
\hline
\rule[-1.2ex]{0mm}{4ex} & & \multicolumn{2}{c}{~$\lambda_S\neq\lambda_g$}  &
 \multicolumn{2}{c|}{~$\lambda_S = \lambda_g$}  \\
 & $\alpha_s(M_Z^2)$  & 0.113   &   0.120  &  0.113  &  0.120    \\
\rule[-1.2ex]{0mm}{4ex}
 & $\Lambda_{\overline{\rm MS}}^{n_f=4}$ (MeV)  & 241   &  344  &  241
  &  344    \\
\hline
    & $(A_g)$          & $ 24.4$ &  $ 14.4$ & $ 2.07$ & $ 0.746$    \\
    & $\lambda_g$      & $-0.41$ &  $-0.51$ & $0.04$ & $0.04$    \\
{\bf Glue}  & $\eta_g$ & $ 6.54$ &  $ 5.51$ & $ 4.56$ & $ 4.38$    \\
    & $\epsilon_g$     & $-4.64$ &  $-4.20$ & $-3.05$ & $-3.85$    \\
    & $\gamma_g$       & $ 6.55$ &  $ 6.47$ & $ 6.83$ & $ 18.1$    \\
\hline
 & $\lambda_u$         & $ 0.60$ &  $ 0.61$ & $ 0.59$ & $ 0.66$    \\
 & $\eta_u$            & $ 3.69$ &  $ 3.54$ & $ 3.67$ & $ 3.55$    \\
 & $\epsilon_{u}$      & $-1.18$ &  $-0.98$ & $-1.02$ & $-1.08$    \\
{\bf Valence}
 & $\gamma_{u}$        & $ 6.18$ &  $ 6.51$ & $ 6.36$ & $ 5.43$    \\
 & $\lambda_d$         & $ 0.24$ &  $ 0.24$ & $ 0.25$ & $ 0.26$    \\
 & $\eta_d$            & $ 4.43$ &  $ 4.21$ & $ 4.45$ & $ 4.18$    \\
 & $\epsilon_d$        & $ 5.63$ &  $ 7.37$ & $ 4.82$ & $ 9.64$    \\
 & $\gamma_d$          & $ 25.5$ &  $ 29.9$ & $ 23.5$ & $ 26.3$    \\
\hline
 & $A_S$               & $ 0.42$ &  $ 0.37$ & $ 0.92$ & $ 0.92$    \\
 & $\lambda_S$         & $ 0.14$ &  $ 0.15$ & $0.04$ & $0.04$    \\
 & $\eta_S$            & $ 9.04$ &  $ 8.27$ & $ 9.38$ & $ 8.93$    \\
{\bf Sea}&$\epsilon_S$ & $ 1.11$ &  $ 1.13$ & $-1.65$ & $-2.34$    \\
 & $\gamma_S$          & $ 15.5$ &  $ 14.4$ & $ 11.8$ & $ 12.0$    \\
 & $A_\Delta$          & $ .039$ &  $ .036$ & $ .040$ & $ .038$    \\
 & $\gamma_\Delta$     & $ 64.9$ &  $ 64.9$ & $ 64.9$ & $ 64.9$    \\
\hline
\end{tabular}                                                                   
\end{center}
\caption{The numerical values of the
starting distribution parameters of the four
sets of partons. Note that $A_g$ is fixed by the momentum sum rule,
and is  therefore not a free parameter.}
\label{tab:paras}
\end{table}
The $\chi^2$ values found in the four fits for
 the deep inelastic structure function subsets of the
data are listed  in Table~\ref{chisquared}.
We have not included the comparisons of each fit with other constraints
such as prompt photon production, $W^\pm $ asymmetry, di-lepton production,
and Drell-Yan $pp/pn$ asymmetry, 
but in every case the quality of the descriptions
is close to that of MRS(A,A$'$) \cite{mrsa,mrsag}.
\begin{table}[tbh]
\begin{center}
\begin{tabular}{|l|c|rrrr|} \hline
 Experiment & \# data    &   \multicolumn{4}{|c|}{ $\chi^2$ }  \\
     &    &  R$_1$ & R$_2$   & R$_3$  & R$_4$  \\ \hline
H1~ $F_2^{ep}$       & 193 & 158 & 149 & 200 &  171 \\
ZEUS~ $F_2^{ep}$     & 204 & 326 & 308 & 378 & 329  \\  \hline
BCDMS~  $F_2^{\mu p}$    & 174 & 265 & 320 & 247 & 311  \\
 NMC~  $F_2^{\mu p}$    & 129 & 155 & 147 & 151 & 152  \\
 NMC~ $F_2^{\mu d}$    & 129 & 139 & 129 & 133 & 132  \\
 NMC~ $F_2^{\mu n}/ 
 F_2^{\mu p}$     & ~85  & 136 & 132 & 136 & 133  \\ 
E665~ $F_2^{\mu p}$    & ~53  &   8 &   8 &   8 &   8  \\ 
SLAC~ $F_2^{e p}$   & 70 & 108 & 95 & 104 & 96      \\ \hline
CCFR~  $F_2^{\nu N}$    & ~66  &  41 &  56 &  42 &  52  \\
CCFR~  $xF_3^{\nu N}$   & ~66  &  51 &  47 &  50 &  44  \\ \hline
 CDF~ $d\sigma/dE_T$ &  & & & & \\
 $50 < E_T < 200$~GeV & 24 & 222 & 52 & 169& 93  \\
 $E_T > 200$~GeV & 11 & 10 & 20 & 9 & 7  \\ \hline
 \Dzero~ $d\sigma/dE_T$ & & & & & \\
 $50 < E_T < 200$~GeV & 14 & 103 & 72 & 50 & 126  \\
 $E_T > 200$~GeV & 12 & 26 & 22 &  12 & 41  \\ \hline
\end{tabular}
\end{center}
\caption{$\chi^2$ values for some of the data used in the global fit.
The $d\sigma / d E_T$ jet data  are predicted, not fitted.}
\label{chisquared}
\end{table}

An idea of the quality of the description of the small $x$ measurements
of $F_2^p$ that is obtained in the R$_1$ and R$_2$ fits can be seen from
the compilation shown in Fig.~2. For comparison we also show the
prediction of the GRV set of (``dynamical") partons \cite{grv}. In the
GRV analysis charm is treated as a heavy quark (rather than as a parton),
this contribution being included in the GRV curves shown.

Fig.~3 displays the $Q^2$ dependence of the exponents $\lambda_g$ and
$\lambda_S$ of the $x^{-\lambda}$  behaviour of the gluon and sea quark
distributions, see Eq.~(\ref{lambdadef}). To be precise the curves are
obtained by assuming that the parton forms (\ref{eq:starting}) describe
the evolved distributions at any $Q^2$ and then determining the
$\lambda_i(Q^2)$ by a five-parameter
$(A_i,\lambda_i,\eta_i,\epsilon_i,\gamma_i)$ fit to 
$xg(x,Q^2)$ and $xS(x,Q^2)$ at each $Q^2$. The upper
plot is obtained from the R$_2$ set of partons, which have $\lambda_S
\neq \lambda_g$  at $Q_0^2 = 1$~GeV$^2$. It shows that the gluon is
valence-like at the input scale. However evolution in $Q^2$ rapidly
steepens the small $x$ shape of the gluon and by $Q^2 \sim 2$~GeV$^2$ it
is already starting  to drive the sea quark distribution, leading to the
DGLAP expectation shown in Eq.~(\ref{eq:dglap}). It is interesting to
note that the cross-over point, $\lambda_g = \lambda_S$, occurs in the
region $Q^2 \sim 5$~GeV$^2$. Thus the behaviour is approximately
compatible with the MRS(A,A$'$) sets of partons which were required to
have $\lambda_g =\lambda_S$ at the $Q_0^2 = 4$~GeV$^2$ starting scale.
The data points shown in Fig.~3 are obtained by the H1 collaboration
\cite{h1} by fitting their measurements of $F_2^p$ for $x < 0.1$ 
in the regions
appropriate to each $Q^2$ bin by the form $x^{-\lambda}$. 
Due to the simplified form of the H1 parametrization and the extensive
$x$ range of the data fitted (which drifts to larger $x$
as $Q^2$ increases),
strictly speaking these values should not be
compared with $\lambda_S$. Nevertheless they do  give a good representation
of the errors as a function of $Q^2$.

We saw in Table~\ref{chisquared} that the $\chi^2$ values for the R$_{2,4}$
descriptions of  the BCDMS $F_2^{\mu p}$ data are significantly
worse than those for R$_{1,3}$. The reason, as already mentioned, is that
it is these data that constrain $\alpha_s(M_Z^2)$ to be $0.113$
in the global fit. The sets with $\alpha_s(M_Z^2) = 0.120$ lead to 
scaling violations for the structure function which are stronger than
those measured by BCDMS. This is illustrated in Fig.~4, which compares
the R$_1$ and R$_2$ descriptions  of the ``medium $x$"
SLAC \cite{slac} and BCDMS \cite{bcdms} proton structure function
measurements. Only those data points
which pass the $W^2 > 10$~GeV$^2$ cut (and which are therefore
included in the global fit) are shown.
The value of $\alpha_s$ is mainly constrained
by the BCDMS data (on both hydrogen and deuterium targets) in the 
$x = 0.35,\; 0.45$ bins. Here
the  preference of the data for the R$_1$ set is evident.
Although the SLAC data have  a slight 
 preference overall for the larger $\alpha_s$
value (see Table~\ref{chisquared}), the error bars are larger and so
the statistical significance is weaker. There is also the possibility that the 
lower $Q^2$ data are contaminated with higher-twist contributions,
which would of course distort the $\alpha_s$ measurement. However, a
careful analysis \cite{mv} of the combined SLAC and BCDMS data, incorporating
a phenomenological higher-twist contribution, yields the value
\begin{equation}
\alpha_s(M_Z^2) = 0.113 \pm 0.005 \; ,
\end{equation}
where the statistical and systematic errors have been combined.

The NLO QCD fits to the medium/large $x$ fixed-target structure function
data are relatively insensitive to the gluon distribution which, in the
MRS global fits, is constrained in this $x$ region by the 
large $p_T$ prompt photon data, see for example Ref.~\cite{mrsag}.
However at small $x$ there is a significant $\alpha_s $ -- gluon correlation,
since $\partial F_2/\partial \ln Q^2 \sim \alpha_s P^{qg} \otimes g$.
This is evident in Fig.~5(a), which shows the gluon distributions of the 
four R fits in the  small $x$ HERA region at $Q^2 = 5$~GeV$^2$.
At small $x$ the R$_1$ gluon is evidently some $10-20\%$ larger
than the R$_2$ gluon.
The same hierarchy is visible in the R$_3$ and R$_4$ gluons, but
these are both slightly steeper. The reason is that constraining
$\lambda_g$ to be equal to $\lambda_S$  forces a larger value
for the former, see Fig.~3, and a correspondingly steeper gluon.
Given that the HERA data clearly favour {\it different}
values for $\lambda_g$ and  $\lambda_S$ (Table~\ref{chisquared})
we may take the spread in the   R$_1$, R$_2$ gluons to
represent the uncertainty in the distribution at small $x$.
It is interesting to note that the previous MRS(A$'$) gluon lies within
this band.

Also shown in Fig.~5(a) is the GRV ``dynamical" gluon \cite{grv}. This
is much larger than any of the four R gluons, and  therefore yields a more
rapid $Q^2$ evolution for the structure function. 
This was already evident in Fig.~2, which  also showed that
such a strong  $Q^2$ dependence is now ruled out by the H1 and ZEUS data.

Figure~5(b) shows the relative behaviour of the
four R gluons at $Q^2 = 10^4$~GeV$^2$, the scale relevant for
the Fermilab jet  data.
As expected, the curves have evolved closer together, although the
qualitative differences seen at $Q^2 = 5$~GeV$^2$ are still apparent.
Quantitatively, we see that the crossover points of the
R$_1$ and R$_3$ gluons (with the same $\alpha_s$) have moved to smaller
$x$ in accordance with the evolution equation,
$\partial g / \partial \ln Q^2 \sim P_{gg} \otimes g$, where $g$ is
sampled in the convolution at slightly larger $x$.

Figure~6 compares the NLO QCD 
predictions\footnote{The calculation uses the next-to-leading-order 
parton level
Monte Carlo JETRAD \cite{jetrad} and the cuts
and jet algorithm applied directly to the partons
are modelled as closely as possible to the experimental
set-up. We thank Nigel Glover for help in performing these calculations.}
 for the single jet inclusive
$E_T$ distribution at the Tevatron,
 using the new R$_1$ and  R$_2$ sets, with data from the
CDF \cite{cdfjet} and \Dzero\ \cite{d0jet} collaborations.
Only the statistical errors on the data are shown,
and in each case the data have been renormalized by a constant
factor to give the best overall fit.\footnote{Note that the normalization
factors shown in Fig.~6 are in each case well within the 
quoted experimental uncertainty.} As the perturbative QCD predictions
become unstable at small $E_T$, only data with $E_T > 50$~GeV are 
included in the comparison. The corresponding
$\chi^2$ values are listed in Table~\ref{chisquared}.
The first and most obvious point to note
is that highest $E_T$ CDF points tend to lie above the theoretical
prediction, a tendency which is {\it not} evident in the \Dzero\ data.
The ``disagreement" with the CDF data  has been the subject of
 much recent discussion, explanations ranging over a variety of new physics
 effects \cite{np} (but see also \cite{mrsjet,cteqjet,kramerjet}).
In view of the fact that the large $E_T$ 
\Dzero\ data appear perfectly compatible
with the R$_{1,2}$ predictions, we believe that it is premature  to
draw any firm conclusions about disagreements with  perturbative
QCD. More interesting is the discriminating power of the high-precision
$E_T \lapproxeq 200$~GeV jet data. Since the $Q^2$ evolution length
is so long, the steepness of the parton distributions in $x$, and therefore
of the jet $E_T$ distribution, is strongly correlated with the 
value of $\alpha_s$ \cite{mrsv}. The effect is clearly
seen in  the differences between the R$_1$ 
($\alpha_s(M_Z^2)=0.113$) 
and R$_2$ ($\alpha_s(M_Z^2)=0.120$) predictions in Fig.~6. One could
therefore, at least in principle, use these data to measure 
$\alpha_s$, see for example Ref.~\cite{ggy}. Such an analysis would
however require a careful consideration of the systematic errors on the
{\it shape} of the measured $E_T$ distribution, and also of
the correlation with the gluon distribution, which is responsible
for a significant fraction of the uncertainty in the
cross section in this $E_T$ range.
For instance, if we consider the R$_1$  and R$_3$ descriptions
(which have the same $\alpha_s$) of the medium-$E_T$ CDF jet data,
the reduction in $\chi^2$ from 222 to 169 in Table~\ref{chisquared}
is due primarily to the difference in slope of the gluons
in the $x \sim 0.1 - 0.2$ region, as shown in Fig.~5(b).
However the slope of the jet cross section is more sensitive
to $\alpha_s$ than to the change in the gluon distribution, as reflected
in the reduction of $\chi^2$ to 52 for these jet data 
for parton set R$_2$.

Taken at face value, the CDF and \Dzero\ jet data would appear to favour
a larger value of $\alpha_s$, of order $0.122$ and $0.118$
respectively. This suggests that  a global analysis incorporating these
data would yield a set of partons very similar  to R$_2$
with $\alpha_s(M_Z^2) \approx 0.120$.

In conclusion, we are able to obtain an excellent and economical
NLO DGLAP-based description of a wide range of deep inelastic
and hard scattering data over an increasingly large kinematic
domain, namely down to $Q^2 = 1.5$~GeV$^2$ and $x \sim 10^{-4}$. The new
HERA data have considerably improved our knowledge of the gluon in the
small $x$ region.
We present\footnote{The {\tt FORTRAN} code for the R sets is available by
electronic mail from W.J.Stirling@durham.ac.uk} two sets of partons
R$_1$ and R$_2$ corresponding to two values of $\alpha_s$,
respectively $\alpha_s(M_Z^2) = 0.113$ (the ``classic"
 DIS value determined
by the scaling violations of the fixed-target data),  and
$\alpha_s(M_Z^2) = 0.120$ (favoured by the HERA $F_2$ and Fermilab
jet data).
It is clear that including the jet data, with statistical errors only, in
the global analysis would in fact discriminate between our two 
sets\footnote{The two sets R$_3$ and R$_4$  with $\lambda_S = \lambda_g$ at
$Q_0^2 = 1$~GeV$^2$ are disfavoured by the HERA data and are used
only for comparison.} of partons in favour of R$_2$.

\section*{Acknowledgements}
We thank Jerry Blazey, Albert De Roeck, Nigel Glover,
Al Goshaw, Mark Lancaster, Robert Thorne, Wu-Ki Tung and Andreas Vogt
for useful discussions.

\newpage
\section*{Figure Captions}
\begin{itemize}
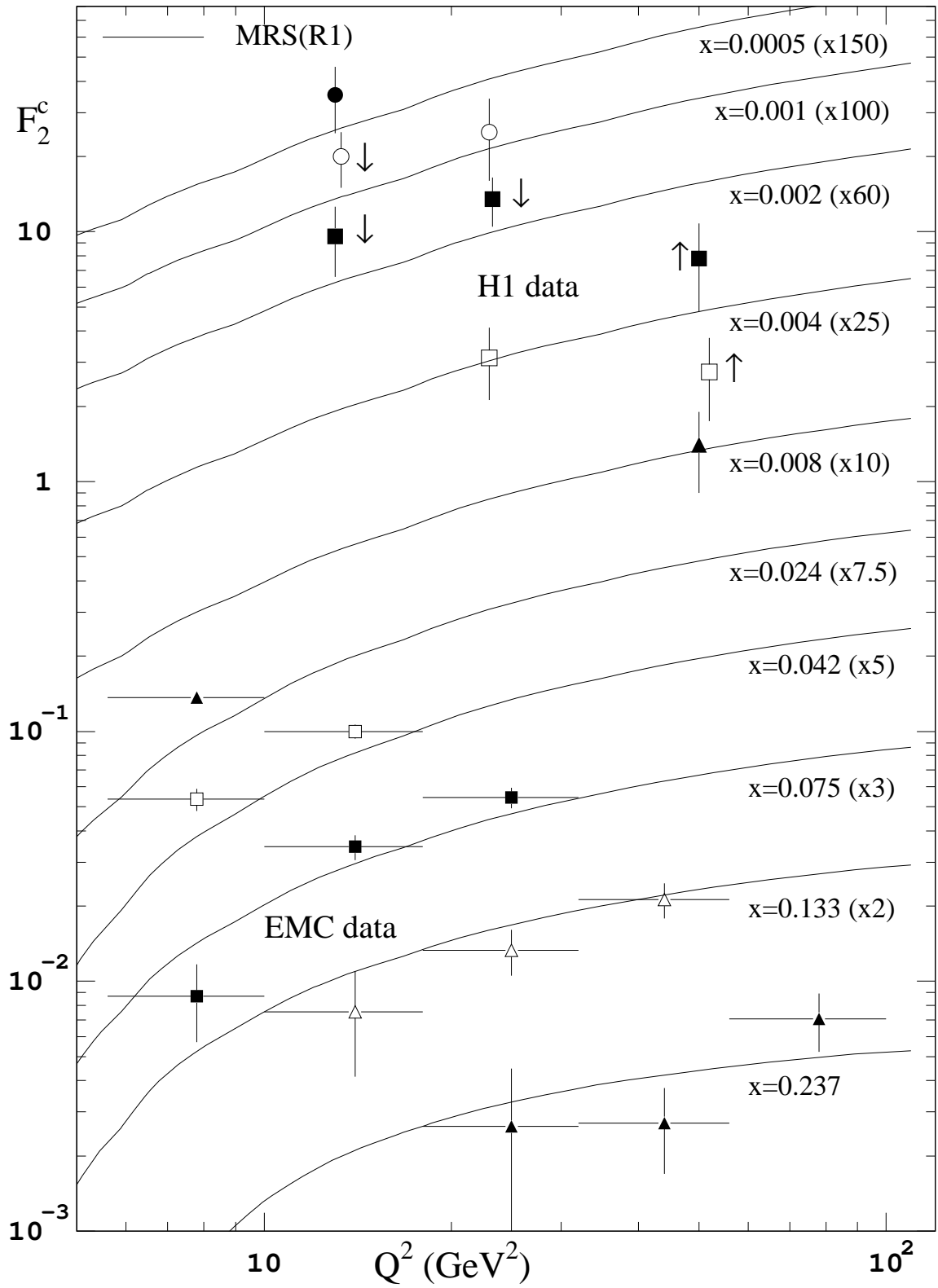
\item[Fig.~1]
The description of the EMC \cite{emc} and
preliminary H1 \cite{h1charm}
data for  $F_2^c(x,Q^2)$ by the R$_1$ set of partons.
The other R$_i$ sets of partons give essentially identical
values of $F_2^c$.

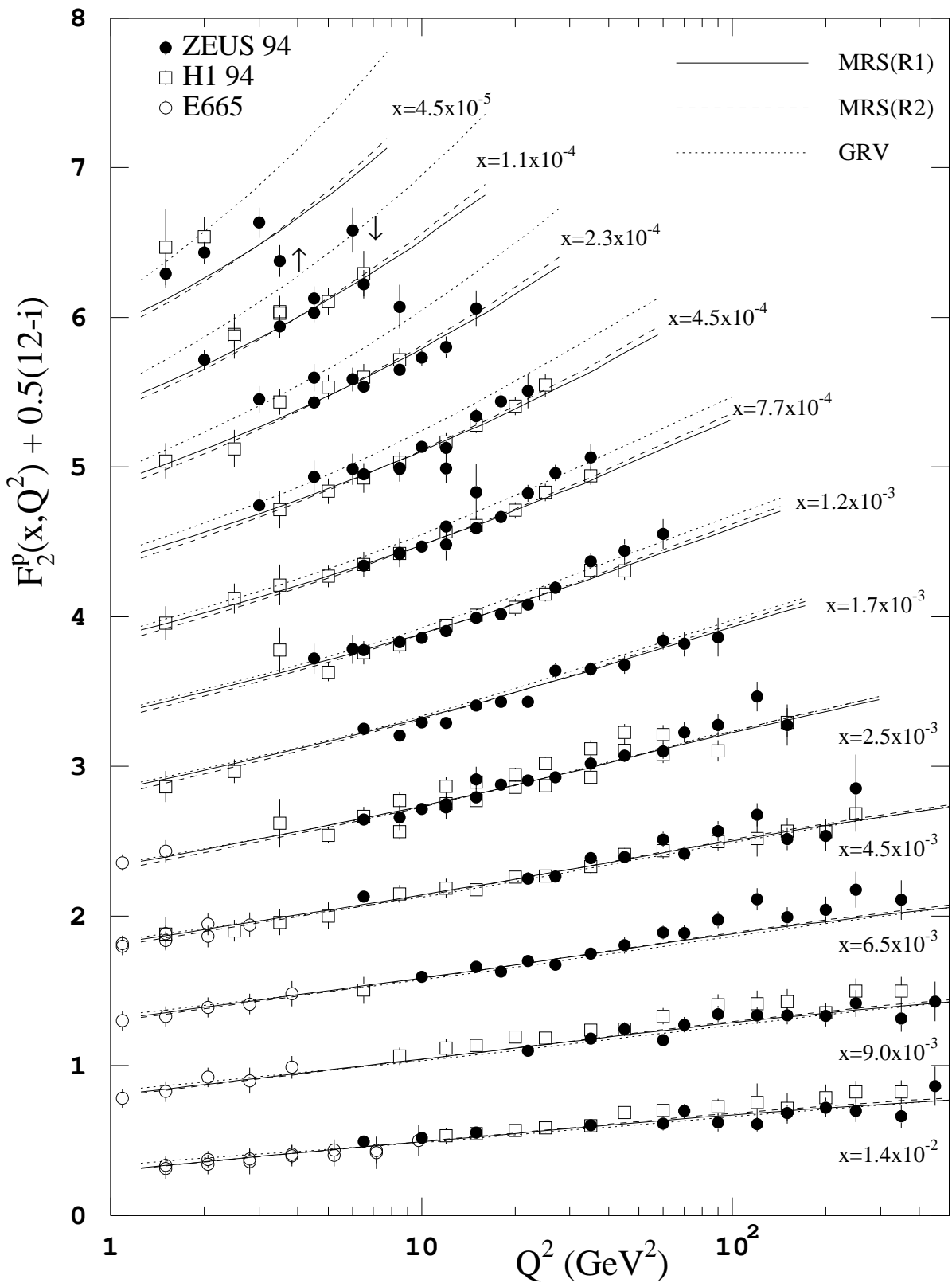
\item[Fig.~2]
The continuous and dashed curves correspond to the values of the proton
structure function $F_2$ obtained from the R$_1$ and R$_2$
sets of partons (which have, respectively, QCD couplings corresponding
to $\alpha_s(M_Z^2) = 0.113$ and $0.120$) at twelve values
of $x$ chosen to be the most appropriate for the new HERA data.
For display purposes we add $0.5(12-i)$ to $F_2$ each time
the value of $x$ is decreased, where $i=1,12$.
For comparison the dotted curves show the prediction obtained from the 
GRV set of partons \cite{grv}.
The experimental data are assigned to the $x$ value which is
closest to the experimental $x$ bin. Thus the ZEUS data\cite{zeus}
are shown in groupings with $x$ values 
$( 3.5,6.3,6.5     \times  10^{-5})$,
$( 1.02,1.20     \times  10^{-4})$,
$( 1.98,2.53      \times  10^{-4})$,
$( 4.0,4.5     \times  10^{-4})$,
$( 6.32,8.00     \times  10^{-4})$,
$(1.02,1.20      \times  10^{-3})$,
$(1.612     \times  10^{-3})$,
$(2.53,2.60      \times  10^{-3})$,
$(4.00     \times  10^{-3})$,
$( 6.325     \times  10^{-3})$,
$(1.02     \times  10^{-2})$,
$(1.612      \times  10^{-2})$,
and the H1 data \cite{h1} in groupings with $x$ values
$( 3.2,5.0     \times  10^{-5})$,
$( 0.80,1.30     \times  10^{-4})$,
$( 2.0,2.5      \times  10^{-4})$,
$( 5.0     \times  10^{-4})$,
$( 6.3,8.0     \times  10^{-4})$,
$(1.3     \times  10^{-3})$,
$(1.585      \times  10^{-3})$,
$(2.0,2.5,3.2      \times  10^{-3})$,
$(3.98,4.0,5.0     \times  10^{-3})$,
$( 6.3     \times  10^{-3})$,
$(8.0     \times  10^{-3})$,
$(1.3      \times  10^{-2})$.
The E665 data \cite{e665}, which are shown on the curves with the
five largest $x$ values, are measured at $x=
 (2.46     \times  10^{-3})$,
$(3.698,5.2    \times  10^{-3})$,
$( 6.934     \times  10^{-3})$,
$(8.933     \times  10^{-3})$,
$(1.225,1.73      \times  10^{-2})$.

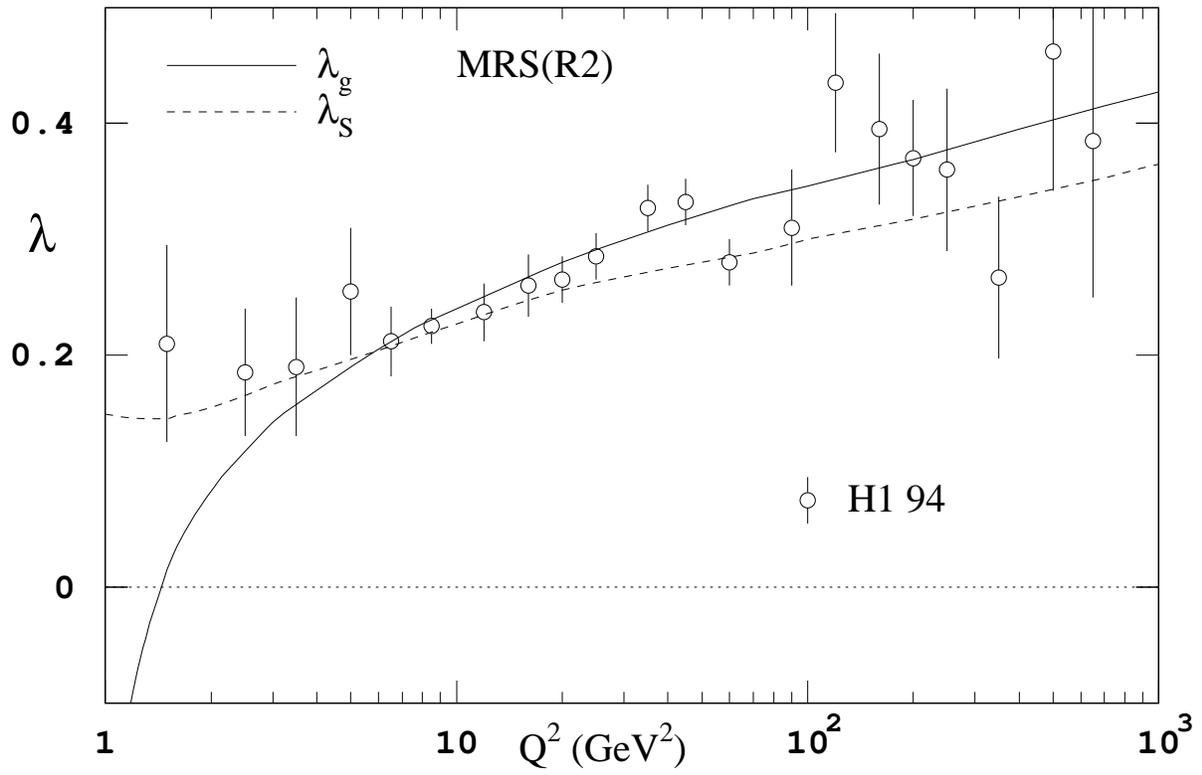
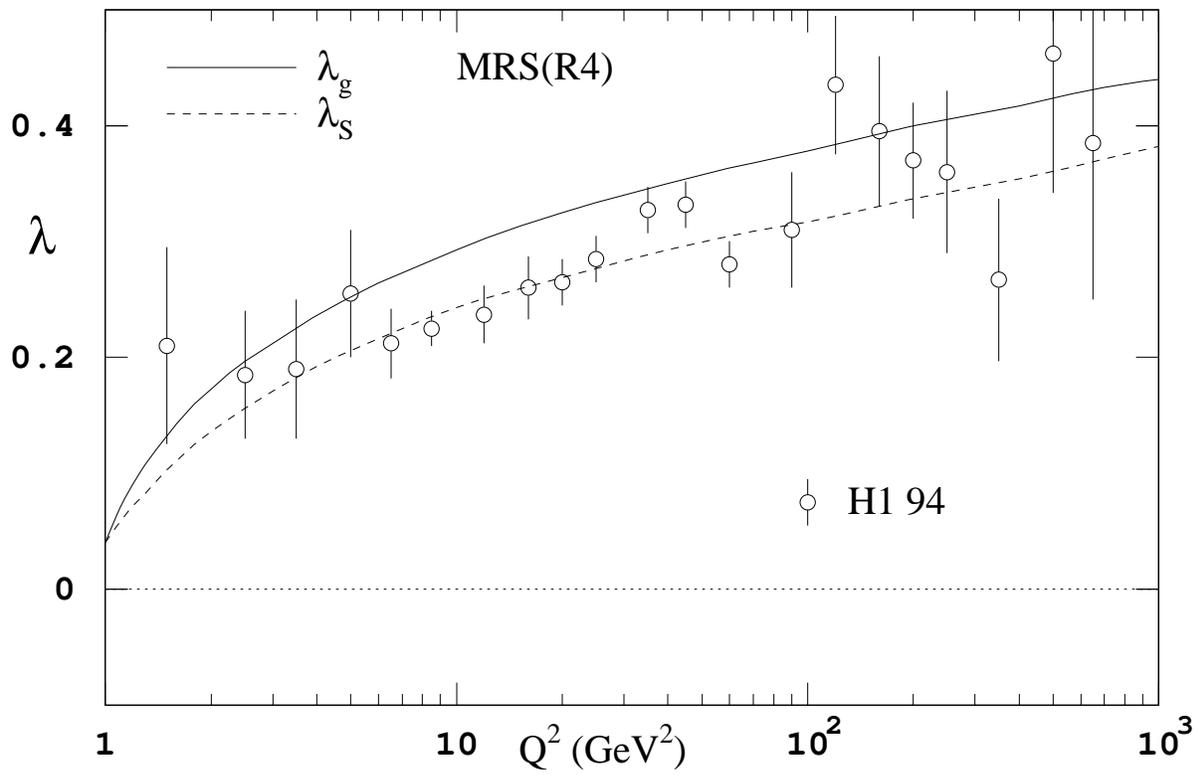
\item[Fig.~3]
The  exponents $\lambda_g$ and $\lambda_S$
as a function of $Q^2$  calculated from the R$_2$ (upper plot) and R$_4$
(lower plot) sets of partons.
Also shown are the
values of $\lambda$ obtained by the H1 collaboration \cite{h1}
by fitting their data  to the form $F_2 = A x^{-\lambda}$ for $x < 0.1$
at different values of $Q^2$.

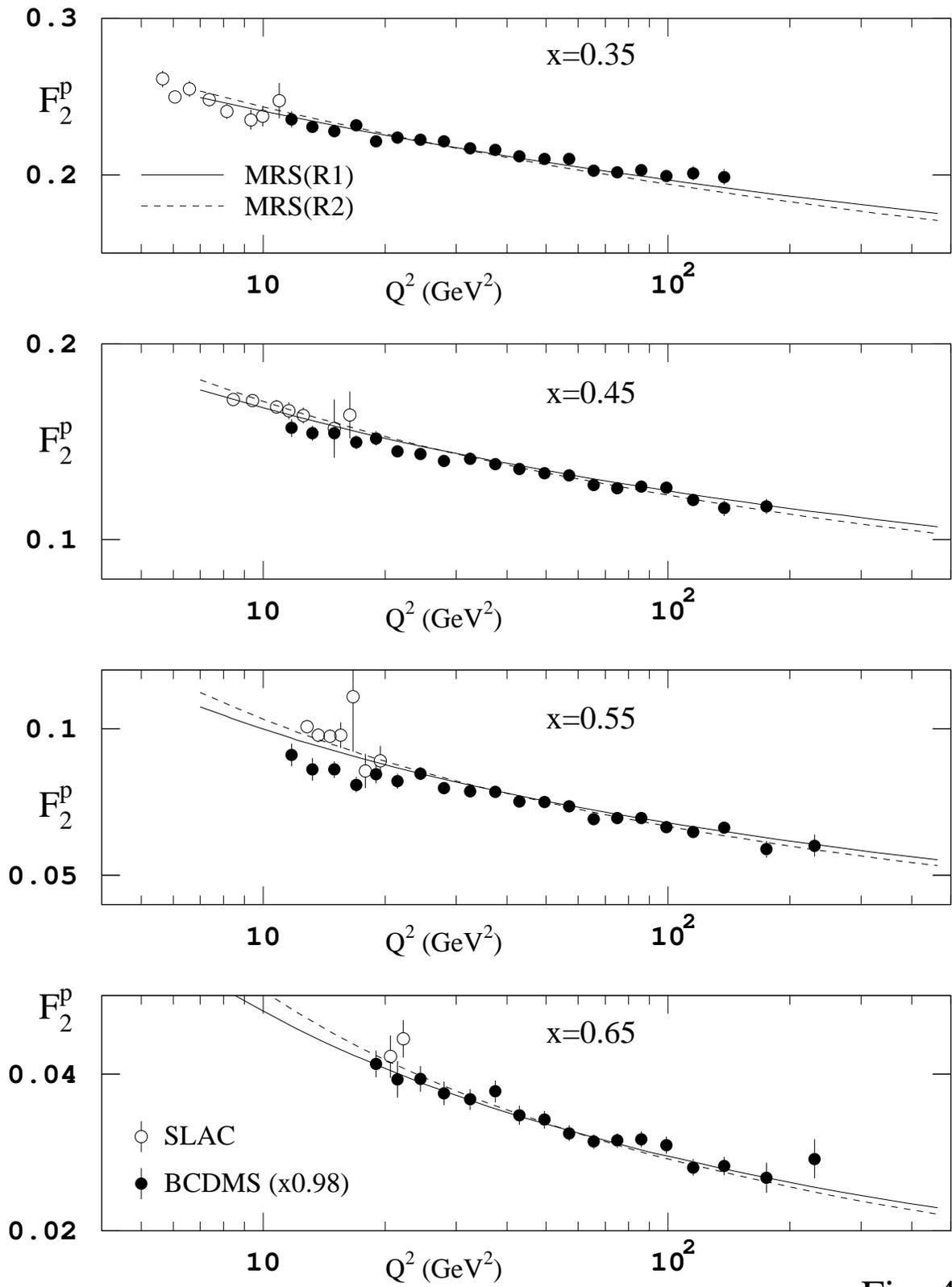
\item[Fig.~4]
The fit to the ``medium" $x$ SLAC \cite{slac} and BCDMS \cite{bcdms}
data for $F_2^p$ by the R$_1$ and R$_2$ sets of partons which have,
respectively, QCD couplings corresponding to
$\alpha_s(M_Z^2) = 0.113$ and $0.120$.

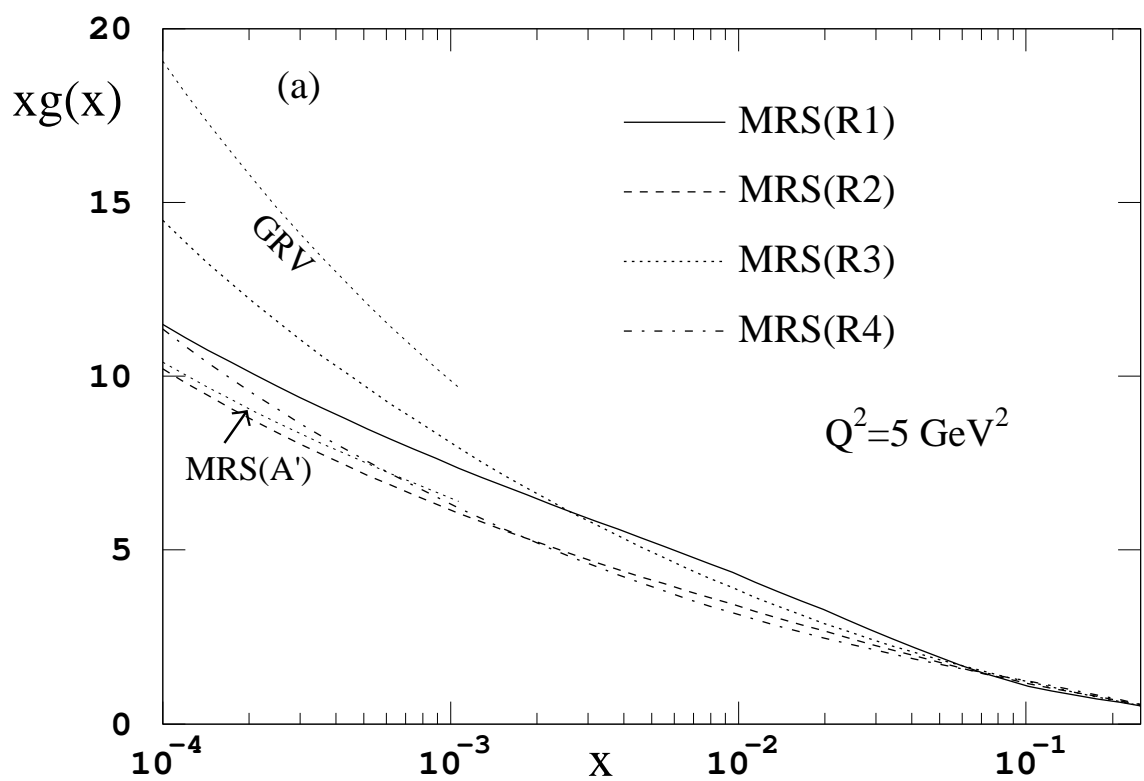
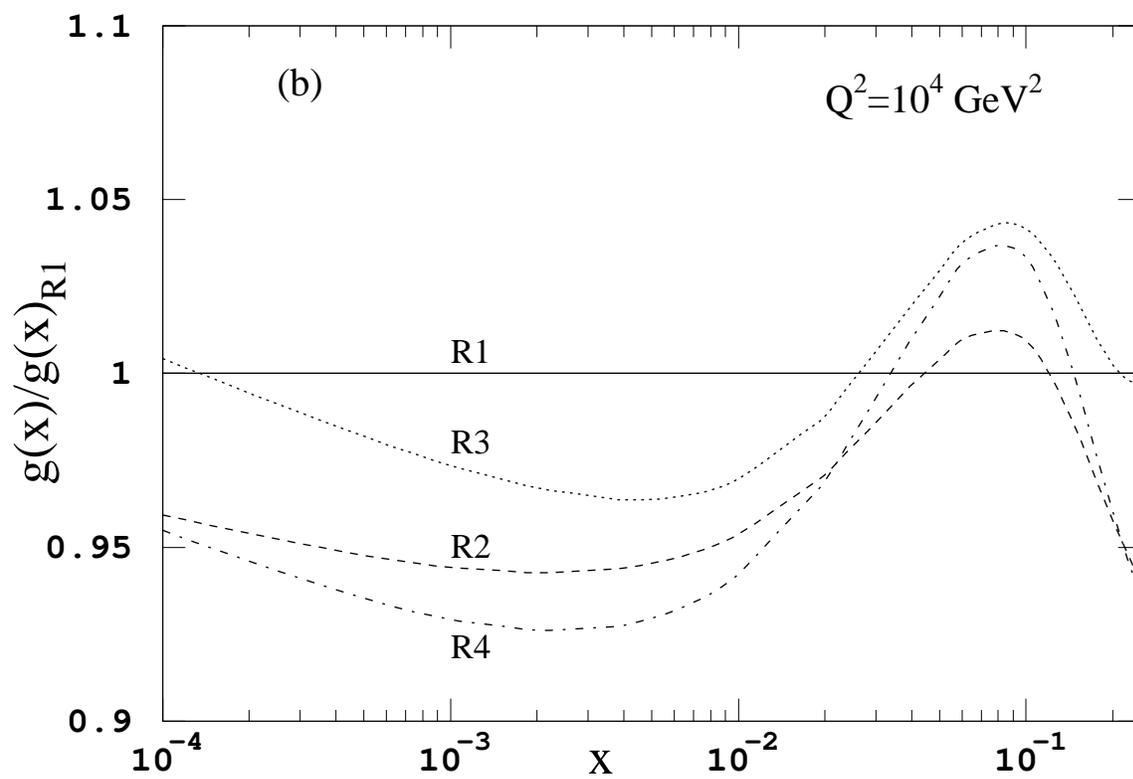
\item[Fig.~5]
The gluon distribution at  (a) $Q^2 = 5$~GeV$^2$
(upper figure) and (b) $10^4$~GeV$^2$ (lower figure). The bold
curves correspond to the new R$_1$, R$_2$, R$_3$ and R$_4$ sets of
partons.
For comparison we also show the MRS(A$'$) \cite{mrsag} and GRV
\cite{grv} gluons for
$x < 10^{-3}$ at $Q^2 = 5$~GeV$^2$.

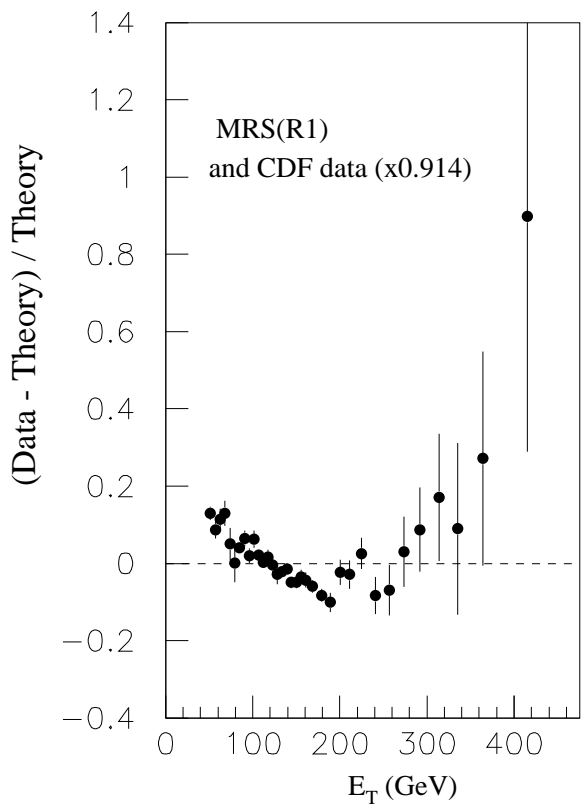
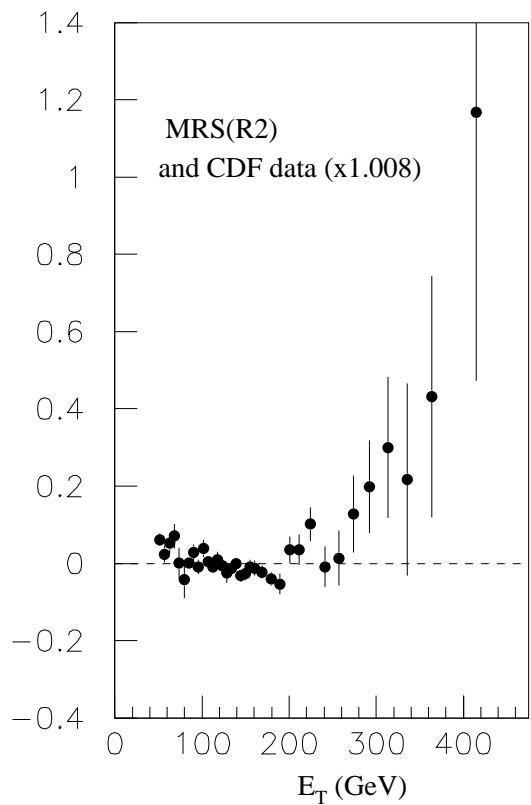
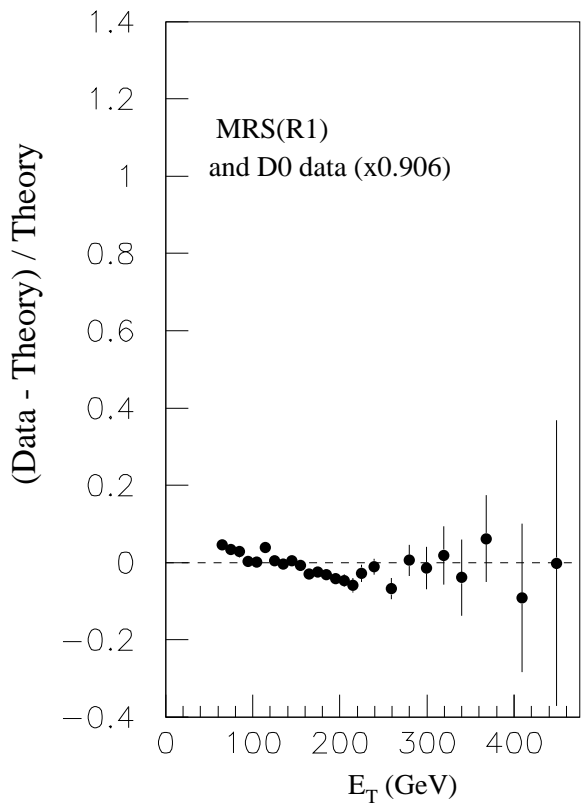
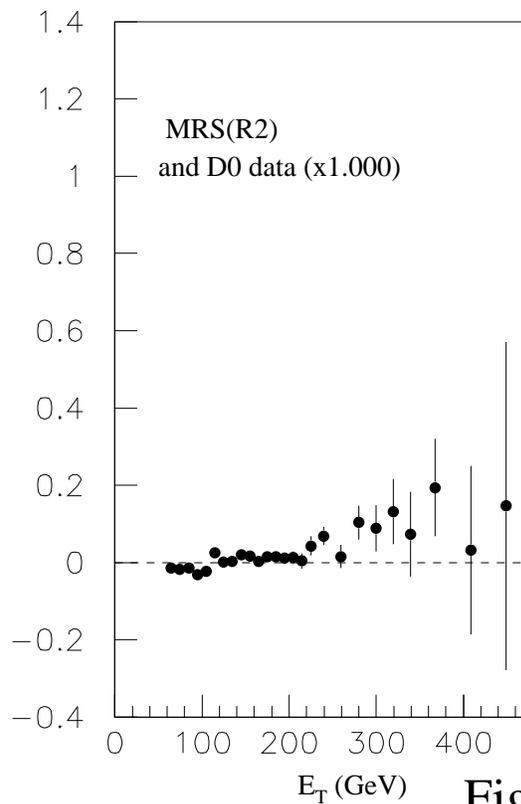
\item[Fig.~6]
The next-to-leading-order QCD description of the CDF \cite{cdfjet}
and \Dzero\ \cite{d0jet} single jet inclusive $E_T$ distribution
by the R$_1$ and R$_2$ sets of partons. These data are not included
in the global analysis. The overall normalization of the QCD predictions
is fitted to the data sets and the value found in each case is shown on
the individual plots.

\end{itemize}
\end{document}